\documentclass[useAMS,twocolumn]{mn2e}
\usepackage{amssymb}
\usepackage{graphicx}

\usepackage[dvips]{color}

\begin{document}

\title[Self-similar solution of magnetized ADAFs with outflow]
{Self-Similar Solution of Hot Accretion Flows with Ordered Magnetic
Field and Outflow}
\author[D.F. Bu, F. Yuan, and F.G. Xie]
{De-Fu Bu$^{1,2,3}$\thanks{E-mail: dfbu@shao.ac.cn (D.F.B.),
fyuan@shao.ac.cn (F.Y.), fgxie@shao.ac.cn (F.G.X.)}, Feng
Yuan$^{1,2}$\footnotemark[1]
and Fu-Guo Xie$^{1,2,3}$\footnotemark[1] \\
$^{1}$Shanghai Astronomical Observatory, Shanghai 200030, China;\\
$^{2}$Joint Institute for Galaxy and Cosmology (JOINGC) of SHAO and USTC;\\
$^{3}$Graduate School of the Chinese Academy of Sciences, Beijing
100039, China; } \maketitle

\begin{abstract}
Observations and numerical magnetohydrodynamic (MHD) simulations
indicate the existence of outflows and ordered large-scale magnetic
fields in the inner region of hot accretion flows. In this paper we
present the self-similar solutions for advection-dominated accretion
flows (ADAFs) with outflows and ordered magnetic fields. Stimulated
by numerical simulations, we assume that the magnetic field has a
strong toroidal component and a vertical component in addition to a
stochastic component. We obtain the self-similar solutions to the
equations describing the magnetized ADAFs, taking into account the
dynamical effects of the outflow. We compare the results with the
canonical ADAFs and find that the dynamical properties of ADAFs such
as radial velocity, angular velocity and temperature can be
significantly changed in the presence of ordered magnetic fields and
outflows. The stronger the magnetic field is, the lower the
temperature of the accretion flow will be, and the faster the flow
rotates. The relevance to observations is briefly discussed.
\end{abstract}


\section{INTRODUCTION}

Advection-dominated accretion flows (hereafter ADAFs) have been
studied extensively (e.g, Narayan \& Yi 1994, 1995; Abramowicz et
al. 1995; see Narayan, Mahadevan \& Quataert 1998 and Kato, Fukue \&
Mineshige 1998 for reviews). It is now rather well established that
this accretion mode exists in the quiescent and hard states of black
hole X-ray binaries and low-luminosity active galactic nuclei (see
Narayan 2005, Yuan 2007, Ho 2008, and Narayan \& McClintock 2008 for
recent reviews). ADAFs only exist below a critical mass accretion
rate. Below this accretion rate, the rate of radiative cooling is so
weak that the viscous heating is balanced by the advection. With the
increase of accretion rate, radiation becomes more and more
important until it becomes equal to the viscous heating at this
critical rate. In this case, the energy advection is equal to zero
(Narayan, Mahadevan \& Quataert 1998). Above this critical accretion
rate, up to another limit close to the Eddington accretion rate,
another hot accretion solution---luminous hot accretion flows
(LHAFs)---was found which is a natural extension of ADAFs (Yuan
2001, 2003). In this solution, the flow is able to remain hot
because the radiative cooling rate, although it is strong, is still
lower than the sum of compression work and viscous heating. Note
that the cool thin solution---standard thin disk---always exists in
the whole range of accretion rates of ADAFs and LHAFs. In the
present work we only focus on the hot accretion flows---ADAFs, but
our discussion should hold for LHAFs. In the early version of ADAFs,
the accretion rate is assumed to be a constant, i.e., there is no
outflow throughout the region. The magnetic field is included, but
only its stochastic component, while the large-scale ordered field
is not considered.

Great developments have been achieved since the discovery of ADAFs.
One is the realization of the existence of outflows in ADAFs, i.e,
only some fraction of the accretion material available at the outer
boundary is actually accreted into the central black hole (Narayan
\& Yi 1994, 1995; Blandford \& Begelman 1999; Stone, Begelman,
Pringle 1999; Igumenshchev \& Abramowicz 1999; Stone \& Pringle
2001). The physical reason for the origin of the outflow is believed
to be that the Bernoulli parameter of the flow is positive (Narayan
\& Yi 1994; Blandford \& Begelman 1999). Another possible mechanism
of the outflow origin is associated with large-scale ordered
magnetic fields, e.g., through the magnetocentrifugal force
(Blandford \& Payne 1982; Henriksen \& Valls-Gabaud 1994; Fiege \&
Henriksen 1996). Observations in the center of our Galaxy supply
strong evidence for the existence of outflows. From $Chandra$
observations combined with Bondi accretion theory, we can predict
the accretion rate at the Bondi radius. Polarization observations at
radio wavebands, however, indicate that the accretion rate at the
innermost region must be significantly smaller than the Bondi value
(Yuan, Quataert \& Narayan 2003). Therefore, a large amount of
material must be lost into outflows.

Another interesting result of numerical magnetohydrodynamic (MHD)
simulations of the hot accretion flow is that a large-scale ordered
magnetic field exists in the inner regions of ADAFs. Independent of
the initial configuration of the magnetic field { \bf(toroidal or
poloidal)} in the main body of the accretion flow the field is
primarily toroidal, with weak radial and vertical components. This
large-scale structure is imposed on the stochastic component of the
magnetic field on small scales (Machida, Hayashi \& Matsumoto 2000;
Hirose et al. 2004).

Both outflows and large-scale magnetic fields can affect the
dynamics of ADAFs significantly. For example, both of them can
effectively transfer angular momentum. These are alternative
mechanisms in addition to the turbulence mechanism associated with
the magnetorotational instability (MRI) proposed by Balbus \& Hawley
(1991; 1998). Stone \& Norman (1994; see also Mouschovias \&
Paleologou 1980) investigate the angular momentum transfer by
magnetic braking effects associated with a large-scale magnetic
field. If the specific internal energy of the outflow is different
from that of the inflow where the outflow originates, the outflow
acts as an extra cooling or heating term in the accretion flow, as
discussed phenomenologically by Blandford \& Begelman (1999). Xie \&
Yuan (2008) parameterize the outflow properties and systematically
investigate the effects of the outflow on the dynamics of the
inflow, in absence of the large-scale magnetic field.

It is thus necessary to investigate the dynamics of ADAFs with
coexistent outflows and large-scale magnetic fields. Several works
have been done recently under the self-similar approximation
(Akizuki \& Fukue 2006; Abbassi, Ghanbari \& Najjar 2008; Zhang \&
Dai 2008) or global solution (Oda et al. 2007). All these works
consider the dynamical effects of the outflow by adopting the form
$\dot{M} \propto r^{s+1/2}$ (e.g., Eq. (\ref{mdot}); Blandford \&
Begelman 1999) to describe the accretion rate while all other
effects such as the probable angular momentum transfer by outflows
are neglected. In Akizuki \& Fukue (2006), Oda et al. (2007), and
Abbassi, Ghanbari \& Najjar (2008), only the toroidal component of
the large-scale magnetic field is considered; thus, the large-scale
magnetic field in their model only supplies an additional force in
the radial direction, while it is unable to transfer angular
momentum. In Zhang \& Dai (2008), although all the three components
of the large-scale magnetic field are included explicitly, their
solutions unfortunately violate the magnetic divergence-free
condition when $s\ne 0$.

In this paper, we investigate the self-similar solutions of ADAFs
with coexistent outflows and large-scale magnetic fields. We assume
the large-scale magnetic field has both $z$ and $\phi$ components,
hence, the large-scale magnetic field does not only affect the
radial force balance of the accretion flow, but also helps to remove
angular momentum. Following Xie \& Yuan (2008), we take into account
the dynamical effects of outflows by considering the differences of
specific angular momentum and specific internal energy between
inflow and outflow. The paper is organized as follows. In $\S2$ we
present the basic MHD equations, which include ordered magnetic
fields and outflows. We address in $\S3$ further assumptions, and
achieve a set of self-similar equations. Numerical results are
presented in $\S4$. We summarize the paper in $\S5$.

\section{BASIC EQUATIONS}

In cylindrical coordinates ($ r, \phi, z$), we investigate the
steady-state, axisymmetric ($\partial/\partial t =
\partial/\partial \phi = 0$) magnetized advection-dominated accretion
flows with outflows. The general MHD equations in Gaussian units
read (e.g., Lovelace, Romanova, \& Newman 1994),
\begin{equation}
\frac{d\rho}{dt}+\rho\nabla\cdot \mathbf{v}=0,\label{cont}
\end{equation}
\begin{equation}
\frac{d\mathbf{v}}{dt}=-\frac{\nabla p}{\rho}-\nabla
\psi+\frac{1}{c}\frac{\mathbf{J}\times\mathbf{B}}{\rho}+
\frac{\mathbf{F}_{vis}}{\rho}, \label{rmon}
\end{equation}
\begin{equation}
\nabla\times\mathbf{B}=\frac{4\pi}{c}\mathbf{J},\end{equation}
\begin{equation}
\nabla\cdot\mathbf{B}=0.\label{diver}\end{equation} The induction
equation of the magnetic field reads,
\begin{equation}
\frac{\partial{\mathbf{B}}}{\partial{t}}=\nabla\times(\mathbf{v}\times
\mathbf{B}-\frac{4\pi}{c}\eta \mathbf{J}).\label{induc}
\end{equation}
The final equation is the energy equation,
\begin{equation}
\rho \left(\frac{d \varepsilon}{d t} -\frac{p}{\rho^2} \frac{d
\rho}{d t}\right)= q_{vis}-q_{rad}\equiv f ~q_{vis}. \label{energy}
\end{equation}
Here $\psi$ is the gravitational potential; $\mathbf{F}_{vis}$ is
the viscous force per unit volume; $\mathbf{B}$ is the magnetic
field; $\eta$ is the magnetic diffusivity; $\mathbf{J}$ is the
current density; $\varepsilon$ is the specific internal energy of
the accretion flow; $q_{vis}$ and $q_{rad}$ are the viscous heating
and radiative cooling rates; $\rho$, $\mathbf{v}$ and $p$ have their
usual meanings. For simplicity, we use the Newtonian potential and
neglect the self-gravity of the accreting gas. The advection factor,
$f$ ($0 \leq f \leq 1$), describes the fraction of the viscous
dissipation energy which is stored in the accretion flow and
advected into the central black hole rather than radiated away.

From the numerical MHD simulations (e.g., Machida, Hayashi \&
Matsumoto 2000; Hirose et al. 2004), it is reasonable to decompose
the magnetic field into one large-scale component and one
fluctuating or turbulent component. The general picture that emerged
from the simulations is that the main body of the accretion flow is
governed by a toroidal magnetic field, especially in its inner
region, while regions near the poles are primarily governed by a
poloidal magnetic field (see Fig. 6 in Hirose et al. 2004), which is
mainly in the vertical direction. Thus, we only consider the $\phi$
and $z$ components ($B_\phi, B_z$) of the large-scale magnetic field
and neglect the radial component ($B_ r$). We describe the effects
of the fluctuating component of the magnetic field in transferring
the angular momentum and in dissipating the energy through the usual
$\alpha$ description. Specifically, the viscous heating rate
$q_{vis}$ in Eq. (\ref{energy}) is associated with the turbulent
component instead of the large-scale ordered component of the
magnetic field. This is because numerical simulations show that the
dissipation mainly comes from the thermalization of the magnetic
energy via magnetic reconnection (Hirose et al. 2006), which is
obviously associated with the turbulent component.

Following Lovelace, Romanova \& Newman (1994), we assume $B_z$ is an
even function of $z$. Furthermore, we neglect the vertical gradient
of $B_z$ in the disc. We take $B_\phi$ to be an odd function of $z$.
We therefore have,
\begin{equation}
\frac{\partial{B_z}}{\partial{z}}=0,
\end{equation}
\begin{equation}
{B_\phi}={B_0}\frac{z}{H}, \quad B_\phi|_{z=H} = -B_\phi|_{z=-H},
\end{equation}
\begin{equation}
{B_0}=B_\phi|_{z=H},
\end{equation}
where $H$ is the half-height of the accretion disc defined as $H
\equiv c_s r /v_K (1+\beta_1)^{1/2}$, where $c_s$ is the isothermal
sound speed, $v_K$ is the Keplerian velocity and $\beta_1$ is
defined below in Eq. (22). It will reduce to its traditional form of
$H = c_s r /v_K$ in absence of the toroidal component ($B_\phi$) of
the ordered magnetic field (Eq. (\ref{zmon2})). Under the above
assumptions, the divergence-free condition of the magnetic field
(Eq. (\ref{diver})) is automatically satisfied.

The specific angular momentum and specific internal energy of the
outflow are in general different from those of the inflow where the
outflow originates. A two-dimensional calculation is ideal but
complicated and out of the scope of our current work. Instead, we
follow the method illustrated in Xie \& Yuan (2008) and use several
parameters ($\xi_1, \xi_2, \xi_3$; see below for definitions) to
characterize the properties of the outflow. In this way, we
integrate the above MHD equations in the vertical direction and get
the following 1.5-dimensional equations to describe the inflow
dynamics (Xie \& Yuan 2008).

The equations of the conservation of mass and momentum are,
\begin{equation}\frac{d\dot{M}(r)}{d r} = \eta_1 4 \pi r \rho v_{z,w},
\label{mass}\end{equation}
\begin{eqnarray}
v_r\frac{d v_r}{d r}+\frac{1}{2\pi r\Sigma}\frac{d\dot{M}(r)}{d r}
\left(v_{r,w}-v_r \right) \nonumber\\
= \frac{{v_\phi}^2}{r}-\frac{GM}{r^2}-\frac{1}{\Sigma}\frac{d(\Sigma
c_s^2)}{d r}-\frac{1}{4\pi\Sigma} \times \nonumber\\
\left[\frac{d}{d r}\left(H B_z^2\right)+\frac{1}{3}\frac{d}{d
r}\left(H B_0^2\right) +\frac{{2}}{3}\frac{{B_
0^2}}{r}H\right],\label{rmon2}
\end{eqnarray}
\begin{eqnarray}
\frac{1}{r} \Sigma v_r\frac{d(r v_\phi)}{d r}+\frac{1}{2\pi
r}\frac{d\dot{M}(r)}{d r}\left(v_{\phi,w}-v_\phi\right) \nonumber\\
=\frac{1}{r^2}\frac{d}{d r}\left(\frac{\alpha\Sigma c_s^2
r^4}{v_K}\frac{d\Omega}{d r}\right)-\frac{B_0 B_
z}{2\pi}.\label{phimon2}
\end{eqnarray}
The static equilibrium in the vertical direction is
achieved between the $z$-component of the gravitational force and
the (gas and magnetic) pressure gradient force,
\begin{equation}
\frac{G M}{r^3}H^2=(1+\beta_1) c_s^2.\label{zmon2}
\end{equation}
The tension term associated with $B_z$ is $B_r d B_z/dr$. It is
neglected since we have assumed $B_r=0$. The energy equation is
\begin{eqnarray}
\frac{v_r}{\gamma-1}\frac{d c_s^2}{dr}-\frac{v_r c_
s^2}{\rho}\frac{d\rho}{d r}+\frac{1}{2\pi
r\Sigma}\frac{d\dot{M}(r)}{d r}(\varepsilon_w-\varepsilon) \nonumber
\\
=f\frac{\alpha c_s^2 r^3}{v_K}\left(\frac{d\Omega}{d
r}\right)^2.\label{energy2}
\end{eqnarray}
Here $v_{r,w}$, $v_{\phi,w}$ and $v_{z,w}$ denote the $r$, $\phi$
and $z$ components of the velocity of the outflow when it is just
launched from the accretion flow (Xie \& Yuan 2008), and
$\varepsilon_ {w}$ and $\varepsilon$ correspond to the specific
internal energy of the outflow and that of the inflow, respectively.
Following Xie \& Yuan (2008), we assume $v_{r,w}=\xi_1 v_r, v_
{\phi,w}=\xi_2 v_\phi$ and $\varepsilon_{w} = \xi_3 \varepsilon$.
Parameter $\gamma$ is the ratio of the specific heats, which goes
from the relativistic value $4/3$ to the typical monatomic gas value
$5/3$. In our calculations we set $\gamma=4/3$. We find our results
are not sensitive to the value of $\gamma$.

The second terms on the left-hand side of Eqs. (\ref{rmon2}) \&
(\ref{phimon2}) are the momenta (radial and angular) taken away from
or deposited into the inflow by the outflow (Xie \& Yuan 2008), and
the third term on the left-hand side of Eq. (\ref{energy2}) is the
internal energy taken away from or deposited into the inflow by the
outflow. If all these $\xi$s are equal to unity, our model will
reduce to those neglecting the discrepancies of specific momenta and
internal energy between the inflow and the outflow (e.g., Akizuki \&
Fukue 2006; Abbassi, Ghanbari \& Najjar 2008; Zhang \& Dai 2008).

Now we discuss the induction equation (Eq. (\ref{induc})). This
equation describes the growth or escape rate of the magnetic field
by dynamo and diffusion. We define the advection rates of the two
components of the magnetic flux as,
\begin{equation}
\dot{\Phi}_{\phi}=\int v_r B_\phi d z,\label{induc1}
\end{equation}
\begin{equation}
\dot{\Phi}_{z}=\int r v_r B_z d z.\label{induc2}
\end{equation}
For a steady-state accretion system considered here, if the dynamo
and the diffusion effects are neglected, the above two quantities
will be constant. However, in a realistic accretion disc, the above
two quantities can vary with radius due to the presence of the
dynamo and the diffusion effects.

\section{SELF-SIMILAR SOLUTIONS}

We seek self-similar solutions in the following forms (e.g., Narayan
\& Yi 1995),
\begin{equation}
v_r=-c_1 ~\alpha v_K,
\end{equation}
\begin{equation}
v_\phi=c_2 ~v_K,
\end{equation}
\begin{equation}
c_s^2=c_3 ~v_K^2.
\end{equation}
We assume the surface density ($\Sigma \equiv 2 \rho H$) to have the
form,
\begin{equation}
\Sigma=\Sigma_0 ~r^s,
\end{equation}
where $\Sigma_0$ and $s$ are two constants, so the accretion rate
takes the form,
\begin{equation}
\dot{M}=2\pi\alpha c_1\Sigma_0\sqrt{G
M}r^{s+\frac{1}{2}}.\label{mdot}
\end{equation}
Note that $s=-1/2$ will lead to solutions without outflows.

We assume the magnetic pressure on the surface of the accretion flow
to be proportional to the gas pressure in the following way,
\begin{equation}
\frac{(B_\phi|_{z=H})^2/8 \pi}{\rho c_s^2}=\beta_1, \label{pmag1}
\end{equation}
\begin{equation}
\frac{(B_z|_{z=H})^2/8 \pi}{\rho c_s^2}=\beta_2. \label{pmag2}
\end{equation}
$\beta_1$ and $\beta_2$ are two parameters. The induction equations
will be satisfied if the dynamo and the diffusion of the magnetic
field satisfy the self-similar conditions required by the advection
rate (Eqs. (\ref{induc1}) \& (\ref{induc2})). Typical values of
$\beta_1$ and $\beta_2$ lie in the range $0.01-1$ in hot accretion
flows (e.g., De Villiers, Hawley \& Krolik 2003; Beckwith et al.
2008), but here we also consider the magnetically-dominated case (
$\beta_1,\beta_2>1$). On one hand, the MHD numerical simulation by
Machida, Nakamura \& Matsumoto (2006) shows that when thermal
instability happens in an ADAF, the thermal pressure rapidly
decreases while the magnetic pressure increases due to the
conservation of magnetic flux. This will result in large $\beta_1$
and $\beta_2$ and forms a magnetically-dominated accretion flow. In
addition, an initially weak large-scale field at the outer region
can be amplified due to the compression of the flow as it is
accreted inward. This can also result in large $\beta_1$ and
$\beta_2$. Of course, in this case, the MRI will be suppressed, but
we still keep $\alpha$ in equations for simplicity and this
treatment will not affect our results significantly for our purpose.

\begin{figure*}
\includegraphics[width=7.5cm]{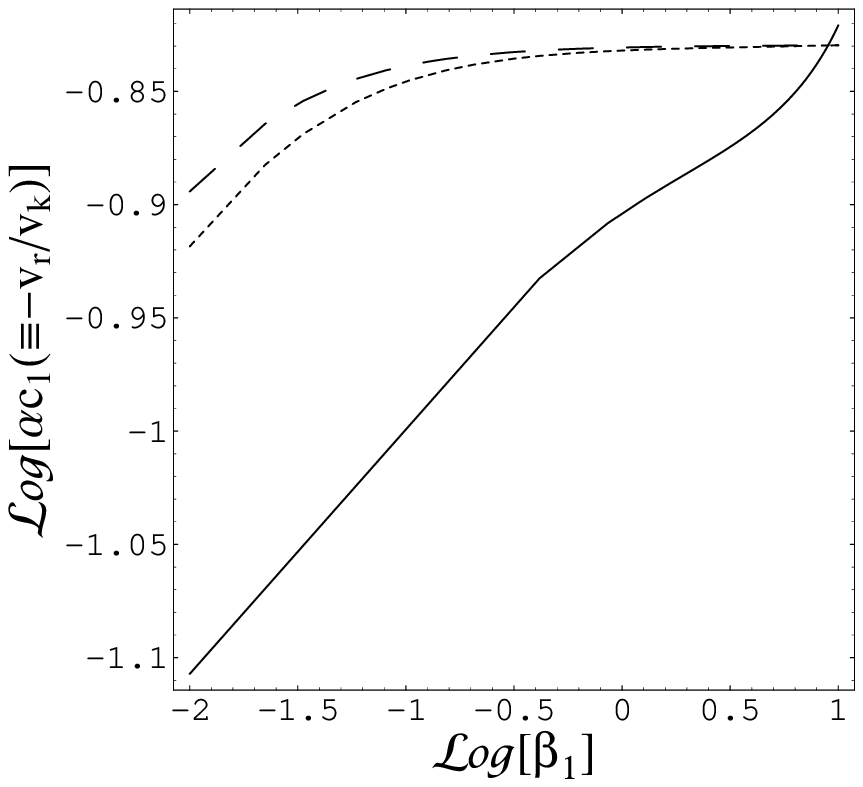}
\includegraphics[width=7.5cm]{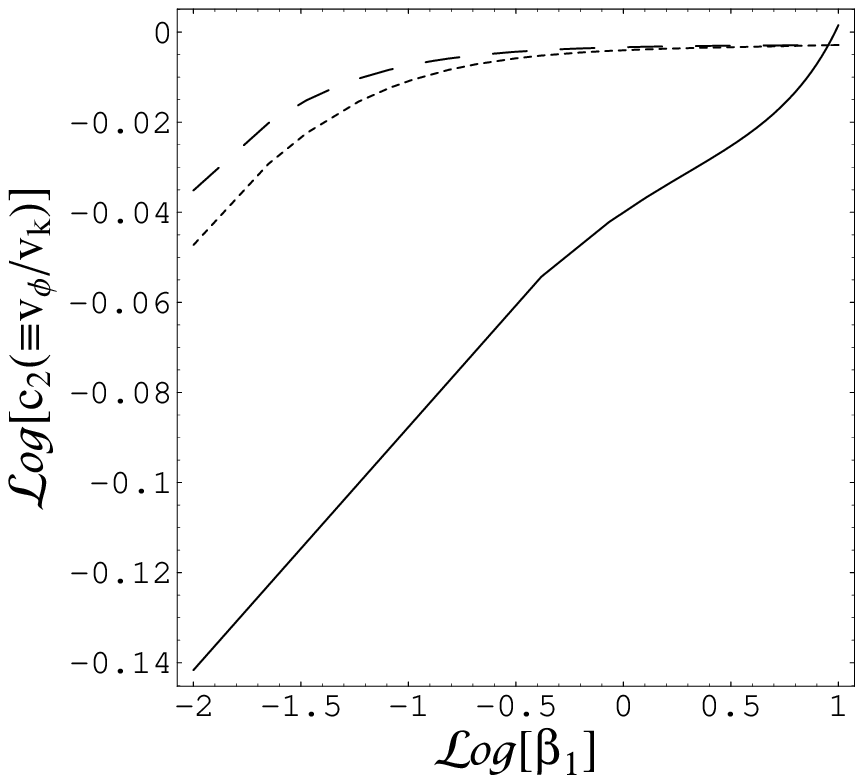}\\
\includegraphics[width=7.5cm]{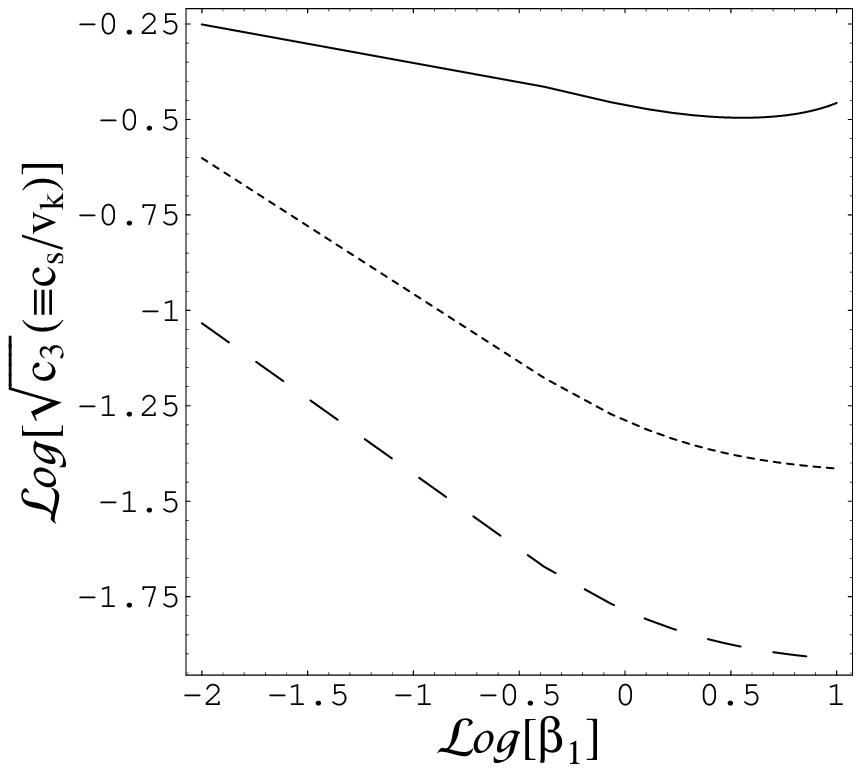}
\includegraphics[width=7.5cm]{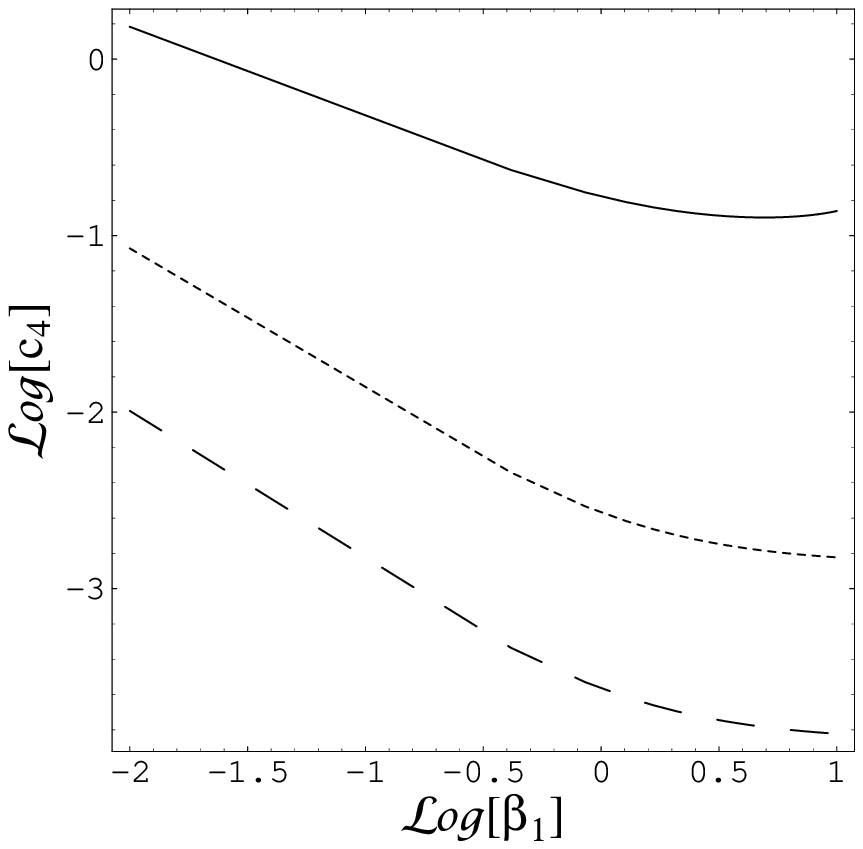}
\caption{The dynamics of the accretion flow in absence of outflows.
Parameters are set as $s=-0.5, \alpha=0.1$ and $f=1$. The solid,
dotted and long-dashed lines correspond to $\beta_2 =0.01, 1$ and
$10$, respectively. The bottom-right panel shows the ratio of the
angular momentum transport due to turbulence (MRI), to that due to
the large-scale magnetic field (see Eq. (\ref{c4}) for definition).}
\label{f1}
\end{figure*}

The momentum conservation and the vertical equilibrium equations
(Eqs. (\ref{rmon2} -- \ref{zmon2})) now reduce to,
\begin{eqnarray}
-\frac{1}{2}c_1^2 \alpha^2-(s+\frac{1}{2})(\xi_1-1)c_1^2
\alpha^2 \nonumber\\
=c_2^2-1-(s-1)c_3-\beta_2 c_3(s-1)\nonumber\\
-\frac{1}{3}\beta_1 c_3(s-1)-\frac{2}{3}\beta_1 c_3,\label{rmon3}
\end{eqnarray}
\begin{eqnarray}
\alpha c_1 c_2-2\alpha (s+\frac{1}{2})(\xi_2-1)c_1 c_2
\nonumber\\
=3\alpha c_2 c_3(1+s)+4\sqrt{c_3}
\sqrt{\frac{\beta_1\beta_2}{1+\beta_1}},\label{phimon3}
\end{eqnarray}
\begin{equation}
H/r=\sqrt{(1+\beta_1)c_3}\label{zmon3}.
\end{equation}
The energy equation becomes,
\begin{equation}
c_1\left[\frac{1}{\gamma-1}+(s-1)+\frac{(s+\frac{1}{2})(\xi_3
-1)}{\gamma-1}\right] =\frac{9}{4}f c_2^2.\label{energy3}
\end{equation}
Obviously for given values of $\alpha, f, s, \beta_1$ and $\beta_2$,
Eqs. (\ref{rmon3}), (\ref{phimon3}) and (\ref{energy3}) form a
closed set of equations of $c_1, c_2$ and $c_3$, which will
determine the dynamics of the accretion flow.

\section{RESULTS}

\begin{figure*}
\includegraphics[width=7.5cm]{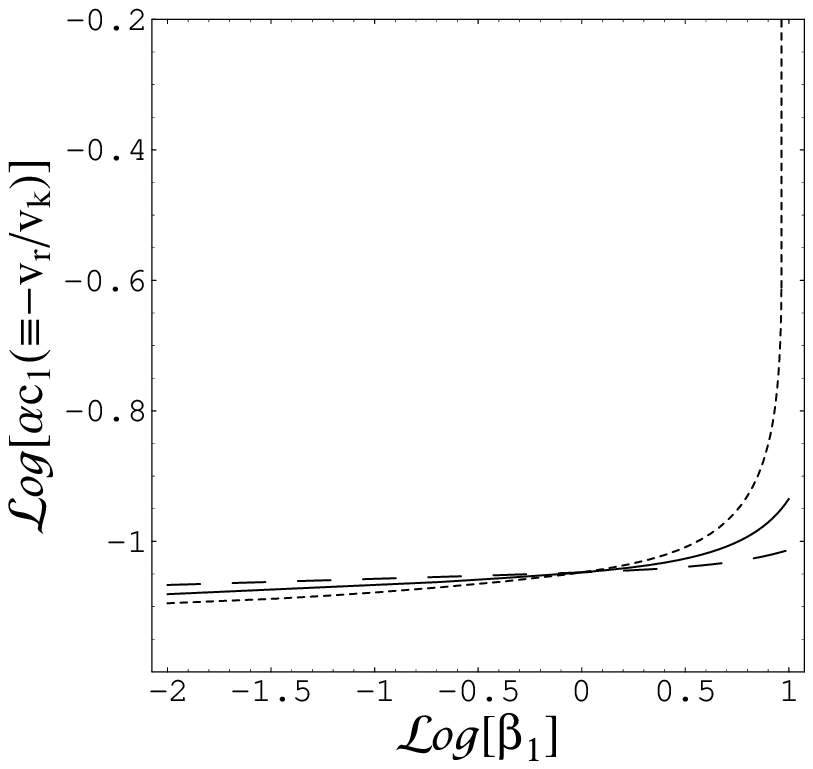}
\includegraphics[width=7.5cm]{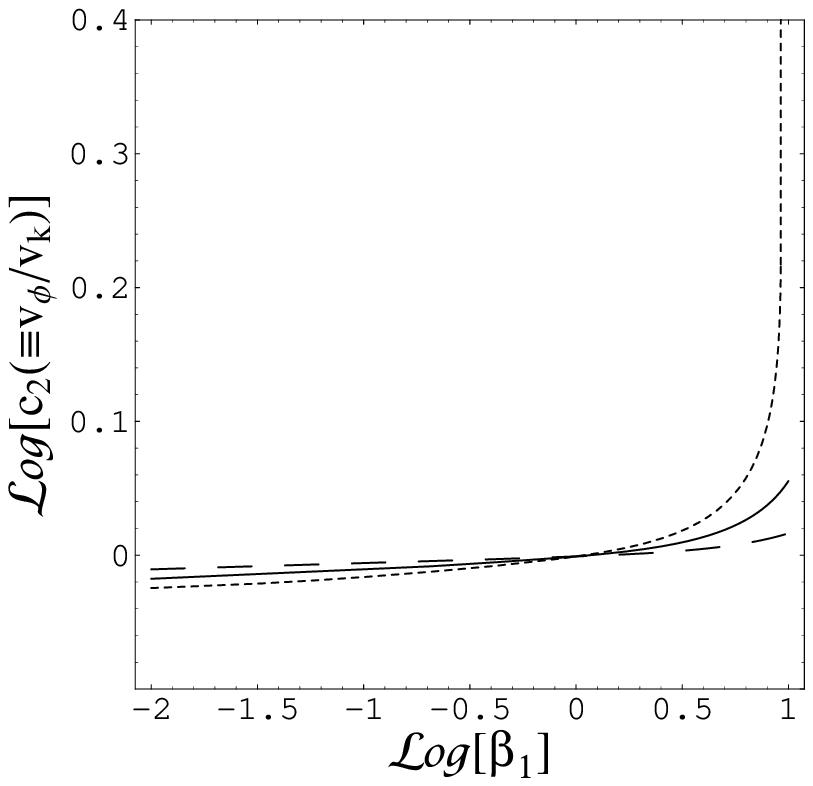}\\
\includegraphics[width=7.5cm]{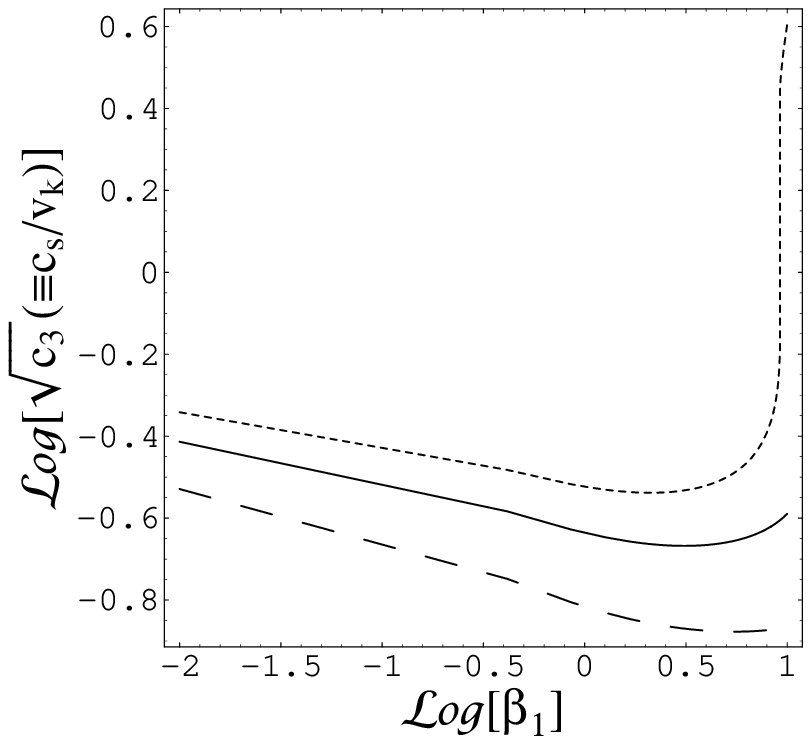}
\includegraphics[width=7.5cm]{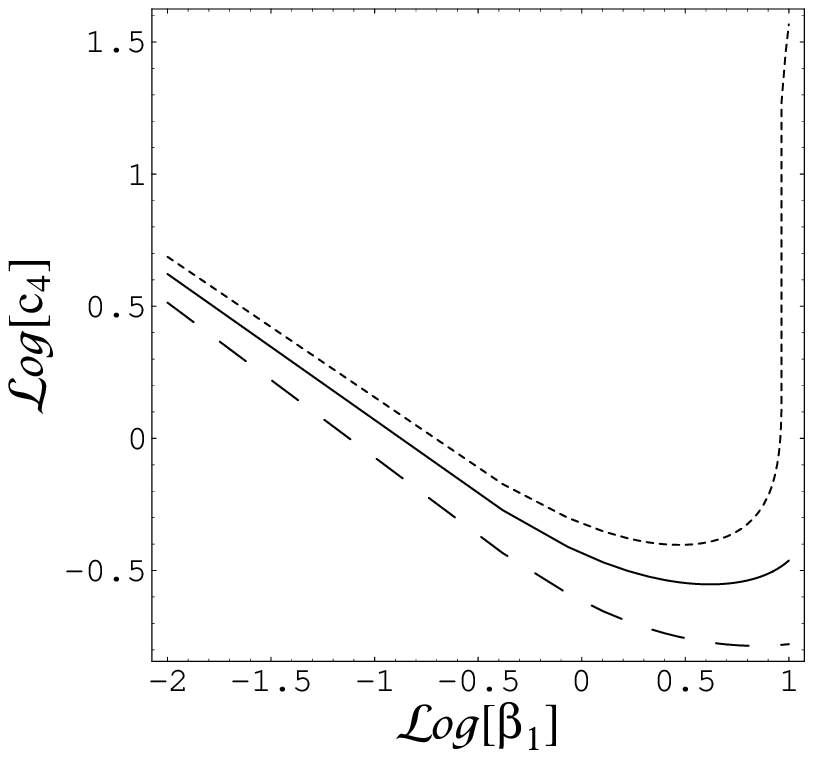}
\caption{The dynamics of the accretion flow when the outflow has
different angular velocity from that of the inflow. Parameters are
$\beta_2=0.01, \alpha=0.1, f=1, s=0.5$ and $\xi_1=\xi_3=1.0$. The
accretion rate $\dot{M}$ is then proportional to $r$. The dotted,
solid and long-dashed lines correspond to $\xi_2 =0.8, 1$ and $1.2$,
respectively. The bottom-right panel shows the ratio ($c_4$; Eq.
(\ref{c4})) of the angular momentum transport due to MRI, to that
due to the ordered magnetic field.} \label{f2}
\end{figure*}

The first and the second terms on the right-hand side of Eq.
(\ref{phimon3}) are angular momentum transfer due to
turbulent-viscosity (MRI) and the large-scale magnetic field,
respectively. We define a new parameter $c_4$ to account for the
ratio of angular momentum transport due to these two mechanisms,
\begin{equation}
c_4=\frac{3 \alpha (1+s) \sqrt{1+\beta_1 } c_2 \sqrt{c_3}}{4
\sqrt{\beta_1 \beta_2}}.\label{c4}
\end{equation}
The large-scale magnetic field will be the dominant mechanism in
angular momentum transport when $c_4$ is below unity.

We first show in Fig. \ref{f1} the effects of the large-scale
magnetic field on the accretion flow. We set $\alpha=0.1, f = 1.0$
and $s=-0.5$ (no outflows). The solid, dotted and long-dashed lines
correspond to $\beta_2 = 0.01, 1.0 $ and $10$, respectively. We can
see from the figure that as $B_\phi$ ($\beta_1$) increases, the
radial ($c_1$) and angular ($c_2$) velocities increase, while the
temperature ($c_3$) and the turbulent-viscosity contribution to
angular momentum transport ($c_4$) decrease. The decrease of $c_4$
is obvious and easy to understand, since the large-scale magnetic
field should be more important in transporting angular momentum as
its strength increases. We find that the magnetic pressure gradient
force (the fourth and fifth terms on the right-hand side of Eq.
(\ref{rmon3})) is a centrifugal force, while the magnetic stress
force (the sixth term on the right-hand side of Eq. (\ref{rmon3}))
is a centripetal force. For the wide range of $\beta_1$ and
$\beta_2$ considered here, the gradient of the total pressure (gas
plus magnetic pressure) serves as a centrifugal force, and it
exceeds the magnetic centripetal force. All these terms are
proportional to the sound speed (or, equivalently, the temperature
$c_3$). The solutions to Eqs. (\ref{rmon3})-(\ref{energy3}), as
shown in Fig. \ref{f1}, indicate that an increase in $B_\phi$
($\beta_1$) results in a decrease in temperature ($c_3$). This in
turn leads to a decrease of the effective centrifugal force, and
subsequently an increase of the infalling velocity $v_r$, as shown
in the upper-left panel of Fig. \ref{f1}. Combined with the energy
equation (Eq. (\ref{energy3})), the angular velocity will increase
correspondingly.

\begin{figure*}
\includegraphics[width=7.5cm]{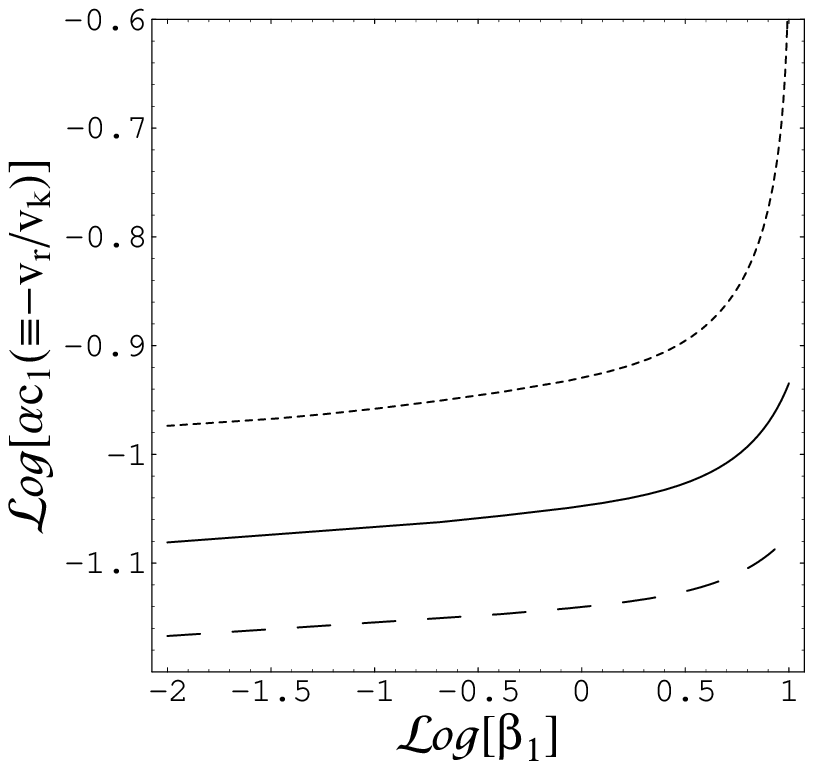}
\includegraphics[width=7.5cm]{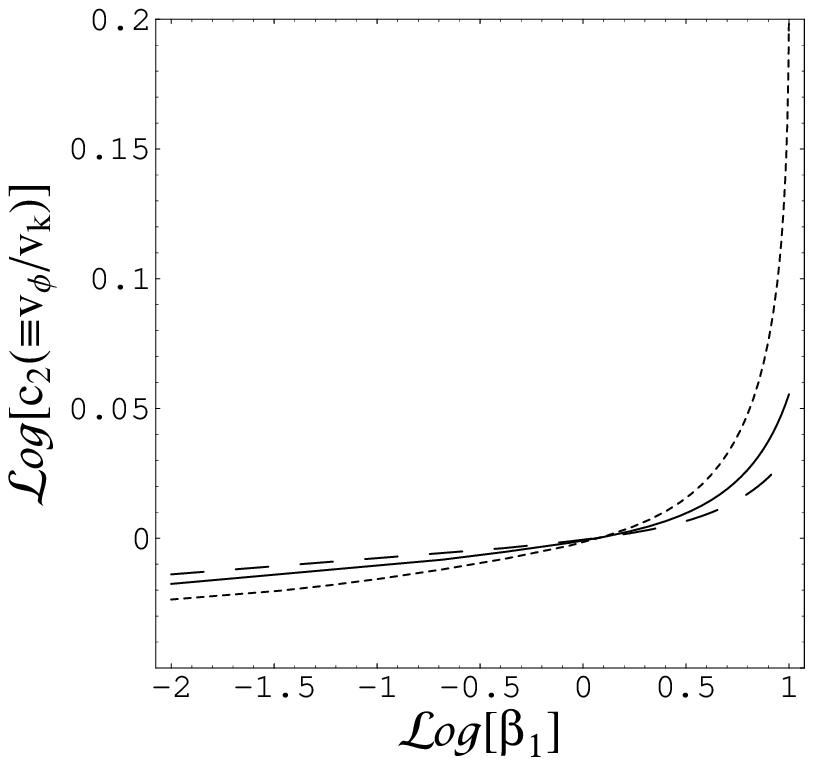}\\
\includegraphics[width=7.5cm]{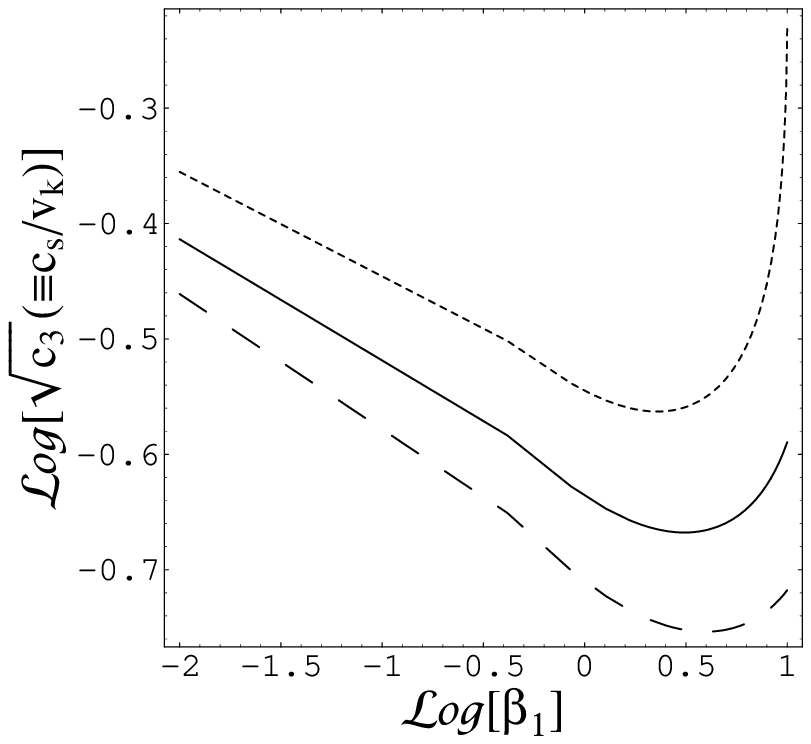}
\includegraphics[width=7.5cm]{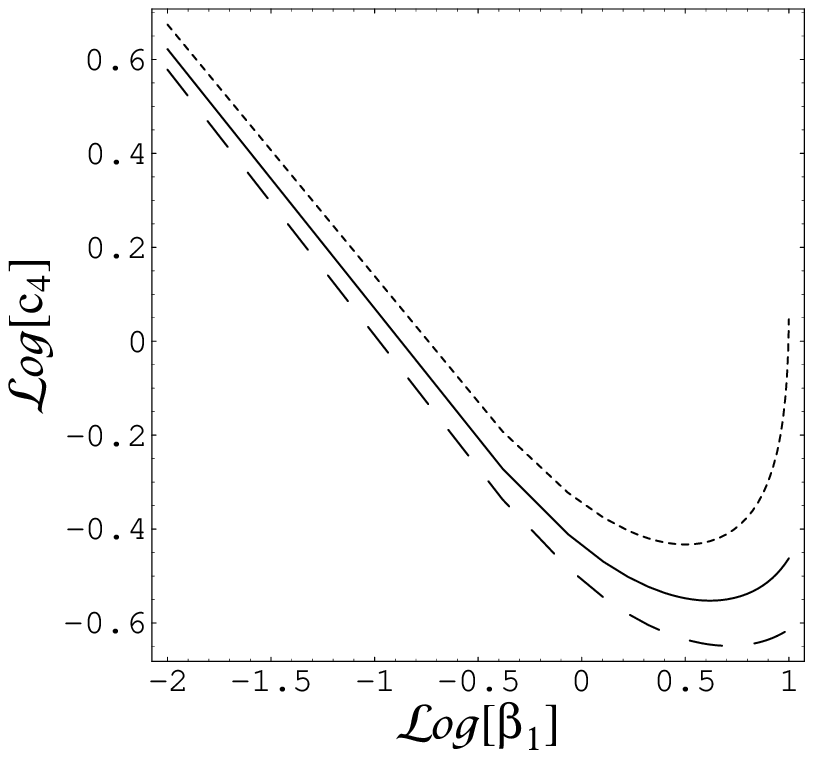}
\caption{The dynamics of the accretion flow when the outflow has
different specific internal energy from that of the inflow.
Parameters are $\alpha=0.1, f=1, s=0.5, \beta_2=0.01$ and $\xi_1 =
\xi_2 = 1$. The dotted, solid, and long-dashed lines correspond to
$\xi_3 =0.8, 1.0$, and $1.2$, respectively. The bottom-right panel
shows the ratio ($c_4$; Eq. (\ref{c4})) of the angular momentum
transport due to MRI, to that due to the ordered magnetic field.}
\label{f3}
\end{figure*}

The change of the $z$ component of the ordered magnetic field has
similar effects. It is evident from Figure \ref{f1} that an increase
in the $z$ component of the magnetic field $B_z$ ($\beta_2$) will
lead to a decrease in temperature ($c_3$). Due to the decrease of
temperature, the effective centrifugal force decreases, so the
infalling velocity increases. From the energy equation (Eq.
(\ref{energy3})), the angular velocity will increase
correspondingly. We thus come to a general conclusion that, the
stronger the magnetic field is, the faster the accretion flow
rotates and falls into the central object, and the lower the
temperature will be (see Figs. (\ref{f2}) \& (\ref{f3}) for
situations with outflows).

The decrease of temperature of ADAFs after taking into account the
large-scale magnetic field may have an important observational
implication. The X-ray spectra of the hard state of black hole X-ray
binaries and some AGNs (such as type 1 Seyfert galaxies) are well
described by a power-law form with a high-energy cutoff. This kind
of spectrum is widely accepted as originating from the thermal
Comptonization process in a hot plasma which is likely to be the
region of a hot accretion flow where most of the radiation
originates (see review by Zdziarski \& Gierli\'nski 2004). The value
of the cutoff energy is determined by the temperature of the
accretion flow, while the slope of the power-law distribution is
determined by the product of temperature and optical depth. From
fitting the observed X-ray spectrum, we can well constrain these two
parameters, namely the optical depth and temperature. On the other
hand, we can also calculate their values from the hot accretion flow
model. Yuan \& Zdziarski (2004) find that while the theoretical
prediction is roughly consistent with observational results, for
some sources the temperature obtained from observation, $\sim 40-60$
keV, is significantly lower than that the lowest value predicted by
the hot accretion flow models, $\ga 80$ keV (see Fig. 1 in Yuan \&
Zdziarski 2004). Our current work indicates that this discrepancy
may be solved by including the large-scale magnetic field (Fig.
\ref{f1}) and the outflow (Figs. \ref{f2}\&\ref{f3}). Unfortunately,
it is difficult to give a quantitative estimation of how strong the
magnetic field is required to be to solve the discrepancy. This is
because our current work is based on the self-similar assumption
which does not hold in the innermost transonic region of the
accretion flow where most of the radiation originates.

Before examining in detail the effects of the outflow on the
dynamics of the inflow, we first check the effects of the
discrepancy of the radial velocity between the outflow and the
inflow (indicated by $\xi_1$). Consistent with Xie \& Yuan (2008),
we find that for relatively weak outflows (e.g., $s < 0$) and
reasonable values of $\xi_1$, the effects are of minor importance.
Thus for the discussions below, we set $\xi_1 = 1$.

We present in Fig. \ref{f2} the effects of angular momentum
discrepancy between the inflow and the outflow. We set $\alpha=0.1,
f=1, \xi_1=\xi_3=1$ and $\beta_2=0.01$. Parameter $s=0.5$ means
$\dot M (r)\propto r$, which corresponds to a relatively strong
outflow (Blandford \& Begelman 1999). The dotted, solid and
long-dashed lines correspond to $\xi_2=0.8, 1$ and $1.2$,
respectively.  Parameter $\xi_2 > 1$ means that the outflow helps to
remove angular momentum from the inflow.

The accretion flow has distinctive properties between $\beta_1 < 1$
and $\beta_1 > 1$. When $\beta_1$ is below unity, the effects of the
large-scale magnetic field can be neglected. As $\xi_2$ increases
from $0.8$ to $1.2$, we can see from Fig. \ref{f2} that the radial
and angular velocities increase while the temperature decreases. We
can understand these results in the following way. From Eq.
(\ref{energy3}), we see that $c_1$ and $c_2$ will increase or
decrease together (note the magnetic terms are negligible). If both
of them would decrease, from Eq. (24) we know that $c_3$ must
increase; while this is in conflict with what Eq. (25) implies if
$\xi_2$ increases. Thus both $c_1$ and $c_2$, i.e., the radial and
angular velocities, must increase when $\xi_2$ increases, and the
temperature decreases. Physically, the increase of the radial
velocity is explained by the decrease in the gas pressure gradient
force with the decreasing temperature. We note that these
self-similar results are consistent with the global solutions of Xie
\& Yuan (2008). We can understand the results for $\beta_1 > 1$ in a
similar way, except that in this case the magnetic force (magnetic
pressure and stress forces) exceeds the gas pressure gradient force.
The critical value of $\beta_1$ corresponds to the balance between
the forces of gas pressure and magnetic pressure, and the value is
related to the strength of the outflow. For cases considered here
($s= 0.5$), this critical value is $1$.

From the upper-right panel of Fig. \ref{f2}, we see that the
accretion flow will be super-Keplerian when the toroidal magnetic
field is very strong (i.e., $\beta_1 > 1$). This is because the
strong magnetic stress force serves as a centripetal force.

Figure \ref{f3} shows the case when the specific internal energy of
the outflow is different from that of the inflow. We take parameters
$\alpha=0.1, f=1, \beta_2=0.01, s=0.5$ and $\xi_1=1$, as before.
Unlike the case shown in Fig. \ref{f2}, here we set $\xi_2 = 1$. The
dotted, solid and long-dashed lines correspond to $\xi_3 = 0.8, 1$
and $1.2$, respectively. The outflow then plays an extra cooling
role for the inflow if its specific internal energy is higher than
that of the inflow ($\xi_3 > 1$), or a heating role otherwise. We
can see from the figure that for the case of a weak magnetic field
($\beta_1<1$), when $\xi_3$ increases from $0.8$ to $1.2$, the
temperature ($c_3$) and radial velocity of the inflow decrease while
the angular velocity increases. The reason is as follows. From Eq.
(\ref{phimon3}) we know $c_1$ and $c_3$ must increase or decrease
together. If both would increase with increasing $\xi_3$, from Eq.
(\ref{rmon3}) we know that $c_2$ must decrease, which is in conflict
with Eq. (\ref{energy3}). The results of a strong magnetic field
($\beta_1>1$) can be understood in a similar way.

Unlike the case without outflows (Fig. \ref{f1}), we find from Figs.
\ref{f2} \& \ref{f3} that, when the toroidal component of the
large-scale magnetic field is strong (e.g., $\beta_1 \approx 10$),
the accretion flow is super-Keplerian. This is because, for the
cases of strong outflows ($s=0.5; \dot{M} \propto r$) considered in
Figs. \ref{f2} \& \ref{f3}, the ``centrifugal force'' (gradient of
the gas plus magnetic pressure) is much weaker than in the case
without outflows (Fig. \ref{f1}); while the ``centripetal force''
(magnetic stress force) does not vary too much.

Finally, we note that in the limits of $f,\beta_1,\beta_2
\rightarrow 0$, $s \rightarrow -0.5$ and $\xi_1,\xi_2,
\xi_3\rightarrow 1$, i.e., the accretion flow is radiatively
efficient, and is devoid of outflows and magnetic fields, our
equations will lead to the solution with $v_r \rightarrow 0 $,
$c_s\rightarrow 0$ and $v_\phi\rightarrow v_K $, which corresponds
to a standard thin disc, as expected.

\section{SUMMARY}

Numerical MHD simulations show that both large-scale ordered
magnetic fields and outflows exist in the inner regions of an ADAF.
In this paper, we have investigated their influences on the dynamics
of the accretion flow in a self-similar approach. We assume that in
addition to the stochastic component, the magnetic field has a
strong large-scale ordered component. The magnetic pressure and
stress forces associated with the large-scale field thus will affect
the dynamics of the accretion flow. We adopt the conventional
$\alpha$ description to mimic the angular momentum and heating
effects associated with the tangled magnetic field. We find that
when the large-scale magnetic field gets stronger, the radial and
angular velocities of the accretion flow increase, while the
temperature decreases. The large-scale field will be more important
than the turbulent-viscosity in transferring angular momentum even
when the field is moderately strong (Fig. \ref{f1}).

We have obtained the self-similar solutions with both large-scale
magnetic fields and outflows. The outflow does not only change the
radial profile of the mass accretion rate, but is also able to take
away angular momentum and energy from the inflow. We parameterize
the physical properties of the outflow, namely its radial velocity,
specific angular momentum and specific internal energy. We find that
when the specific angular momentum of the outflow increases, the
temperature of the inflow decreases, while the radial and angular
velocities increase (decrease) when the magnetic field is weak
(strong) (Fig. \ref{f2}). When the specific internal energy of the
outflow increases, the temperature and radial velocity of the inflow
decrease, while the angular velocity increases (decreases) when the
large-scale magnetic field is weak (strong) (Fig. \ref{f3}). When
both outflows and large-scale magnetic fields are present, we find
that the inflow could be super-Keplerian if the field is very
strong.

The decrease in temperature of the inflow in presence of large-scale
magnetic fields and/or outflows could explain a puzzle that the
predicted temperature of hot accretion flows is higher than that
obtained from fitting the observational data.

\section{ACKNOWLEDGMENTS}

We are grateful for a discussion with Ramesh Narayan. We thank the
anonymous referee for his/her helpful suggestions and careful
reading, which improve the manuscript. This work was supported in
part by the Natural Science Foundation of China (grants 10773024,
10833002, 10821302, and 10825314), One-Hundred-Talent Program of
Chinese Academy of Sciences, and the National Basic Research Program
of China (2009CB824800).


\begin{thebibliography}{}
\bibitem[]{} Abbassi, S., Ghanbari, J., \& Najjar, S. 2008,
MNRAS, 388, 663
\bibitem[]{} Abramowicz, M. A., Chen, X., Kato, S., Lasota J.-P.,
\& Regev, O. 1995, ApJ, 438, L37
\bibitem[]{} Akizuki, C., \& Fukue, J. 2006, PASJ, 58, 469
\bibitem[]{} Blandford, R. D., \& Begelman, M. C. 1999, MNRAS, 303, L1
\bibitem[]{} Blandford, R. D., \& Payne, D. G. 1982, MNRAS, 199,
883
\bibitem[]{} Balbus, S., \& Hawley, J. F. 1991, ApJ, 376, 214
\bibitem[]{} Balbus, S., \& Hawley, J. F. 1998, RvMP, 70, 1
\bibitem[]{} Beckwith, K., Hawley, J. F., \& Krolik, J. H. 2008,
ApJ, 678, 1180
\bibitem[]{} De Villiers, J.-P., Hawley, J.~F., \& Krolik, J.~H. 2003,
ApJ, 599, 1238
\bibitem[]{} Fiege, J. D., \& Henriksen, R. N. 1996, MNRAS, 281,
1005
\bibitem[]{} Hirose, S., Krolik, J. H., De Villiers, J. P., \&
Hawley, J. F. 2004, ApJ, 606, 1083
\bibitem[]{} Hirose, S., Krolik, J. \& Stone, J. 2006, ApJ, 640, 901
\bibitem[]{} Henriksen, R. N., \& Valls-Gabaud, D. 1994, MNRAS,
266, 681
\bibitem[]{} Ho, L. 2008, ARA\&A, in press (arxiv:0803.2268)
\bibitem[]{} Igumenshchev, I.~V. \& Abramowicz, M.~A. 1999, MNRAS, 303, 309
\bibitem[]{} Kato, S., FuKue, J., \& Mineshige, S., 1998,
Black Hole Accretion Disks, Kyoto Univ. Press, Kyoto
\bibitem[]{} Lovelace, R. V. E., Romanova, M. M., \& Newman, W.
I. 1994, ApJ, 437, 136L
\bibitem[]{} Machida, M., Hayashi, M.~R., \& Matsumoto, R. 2000, ApJ, 532, L67
\bibitem[]{} Machida, M., Nakamura, K.~E., \& Matsumoto, R. 2006, PASJ, 58, 193
\bibitem[]{} Mouschovias, T. Ch., \& Paleologou, E. V. 1980, ApJ,
237, 877
\bibitem[]{} Narayan, R. 2005, ApSS, 300, 177
\bibitem[]{} Narayan, R., Mahadevan, R., \& Quataert, E. 1998, in
``Theory of Black Hole Accertion Discs'', eds. M.A. Abramowicz, G.
Bjornsson, J. E. Pringle, Cambridge Univ. Press, Cambridge, p. 148
\bibitem[]{} Narayan, R., \& McClintock, J. E. 2008, in ``Jean-Pierre Lasota,
X-ray binaries, accretion disks and compact stars'' New Astronomy
Reviews, eds. M.A. Abramowicz and O. Straub (Elsevier, 2008).
arxiv:0803.0322
\bibitem[]{} Narayan, R., \& Yi, I. 1994, ApJ, 428, L13
\bibitem[]{} Narayan, R., \& Yi, I. 1995, ApJ, 452, 710
\bibitem[]{} Oda, H., Machida, M., Nakamura, K. E., \& Matsumoto, R. 2007, PASJ,
59, 457
\bibitem[]{} Stone, J. M., \& Norman, M. L. 1994, ApJ, 433,
746
\bibitem[]{} Stone, J. M., \& Pringle, J. E. 2001, MNRAS, 322,
461
\bibitem[]{} Stone, J. M., Pringle, J. E., \& Begelman, M. C.
1999, MNRAS, 310, 1002
\bibitem[]{} Xie, F. G., \& Yuan, F. 2008, ApJ, 681, 499
\bibitem[]{} Yuan, F. 2001, MNRAS, 324, 119
\bibitem[]{} Yuan, F. 2003, ApJ, 594, L99
\bibitem[]{} Yuan, F. 2007, in ``The Central Engine of Active Galactic Nuclei'',
ASP Conference Series, Vol. 373, eds. Luis C. Ho and Jian-Min Wang,
p.95
\bibitem[]{} Yuan, F., Quataert, E., \& Narayan, R. 2003, ApJ, 598, 301
\bibitem[]{} Yuan, F., \& Zdziarski, A.~A. 2004, MNRAS, 354, 953
\bibitem[]{} Zdziarski, A.~A., \& Gierli\'nski, M. 2004, PThPS, 155, 99
\bibitem[]{} Zhang, D., \& Dai, Z. G. 2008, MNRAS, 388, 1409
\end{thebibliography}
\end{document}